\documentclass[twocolumn,secnumarabic,amssymb,nobibnotes,aps,prl]{revtex4-1}
\usepackage{color}
\usepackage{graphicx}
\begin{document}

\title{Magnetically-driven electronic phase separation in the  semimetallic ferromagnet EuB$_6$}%

\author{Pintu Das}%
\affiliation{Institute of Physics, Goethe-University Frankfurt, 60438 Frankfurt (M), Germany}

\author{Adham Amyan}%
\affiliation{Institute of Physics, Goethe-University Frankfurt, 60438 Frankfurt (M), Germany}

\author{Jens Brandenburg}%
\affiliation{Institute of Physics, Goethe-University Frankfurt, 60438 Frankfurt (M), Germany}

\author{Peng Xiong}%
\author{Stephan von Moln\'{a}r}%
\affiliation{Department of Physics, Florida State University, Tallahassee, Florida 32306, USA}

\author{Zachary Fisk}%
\affiliation{Department of Physics, University of California, Irvine, California 92697, USA}

\author{Jens M\"uller}%
\affiliation{Institute of Physics, Goethe-University Frankfurt, 60438 Frankfurt (M), Germany}

\date{\today}

\begin{abstract}
From measurements of fluctuation spectroscopy and weak nonlinear transport on the semimetallic ferromagnet EuB$_6$
we find direct evidence for magnetically-driven electronic phase separation consistent with the picture of percolation of magnetic polarons (MP), which form highly conducting magnetically-ordered clusters in a paramagnetic and 'poorly conducting' background. These different parts of the conducting network are probed separately by the noise spectroscopy/nonlinear transport and the conventional linear resistivity.
We suggest a comprehensive and 'universal' scenario for the MP percolation, which occurs at a critical magnetization either induced by ferromagnetic order at zero field or externally applied magnetic fields in the paramagentic region.

\end{abstract}
\maketitle

The observation that the magnetic state of a system critically affects its electronic transport properties is at the heart of spintronics research. Fundamentals of this effect can be studied in a number of more or less complex model systems showing large negative (or even colossal) magnetoresistance (MR) behavior, e.g.\ magnetic semiconductors and mixed-valence perovskites. These materials often have rich phase diagrams, in which many phases exhibit intrinsic, i.e.\ non-chemical, electronic phase separation. Nanoscale phase separation in turn is thought to play a critical role in the emergence of colossal MR (CMR) in such materials \cite{Dagotto2001}.
Therefore, electronic phase separation has been a subject of intensive recent theoretical and experimental interest.
For the present study, we have chosen the low-carrier density ferromagnetic (FM) semimetal EuB$_6$. EuB$_6$ has a cubic lattice symmetry and is magnetically isotropic (Eu$^{2+}$ localized 4f spins in a $^8S_{7/2}$ Hund's rule ground state), yet it shows interesting physics, where the interplay of metallicity and the formation of clustered magnetic phases can be studied on the fundamental level in a 'clean' system.

Despite the simple lattice and magnetic structure, the physical properties at low temperatures, in particular the mechanism of FM ordering in EuB$_6$ and its interplay with CMR behavior is far from being fully understood. While the material undergoes the paramagnetic (PM) to FM transition, it exhibits two anomalous features at $T_{c_1} \sim 15.5$\,K and $T_{c_2} \sim 12.5$\,K in electronic transport and specific heat measurements, which initially had been interpreted as different kinds of FM ordering \cite{DegiorgiPRL1997,CooleyPRB1997}. Applying small magnetic fields drastically suppresses the resistivity at the higher transition $T_{c_1}$.
Different mechanisms have been discussed in order to explain such a CMR effect in nonmanganite systems, e.g.\ the suppression of critical magnetic fluctuations with externally applied magnetic fields \cite{MajumdarPRL1998} or a delocalization of carriers due to the overlap of magnetic polarons (MP) \cite{TeresaNature1997,NyhusPRB1997,SuellowPRB2000}. MP, which first have been suggested in an experimental study of the magnetic semiconductors Eu$_{1-x}$Gd$_x$Se in 1967 \cite{Molnar1967}, are formed when it is energetically favorable for the charge carriers to spin polarize the local moments over a finite distance, i.e.\ the localization length of the charge carriers \cite{KasuyaRMP1968}.
The size of the polaronic clusters
is thus determined by the balance of the increase in kinetic energy of the charge carriers due to their localization and the reduction of exchange energy due to alignment of the local moments \cite{vonMolnarHandbook}.
From small angle neutron scattering experiments, De Teresa \textit{et al.}\ have demonstrated the existence of such clustered phases above the FM ordering temperature $T_C$ of manganite materials \cite{TeresaNature1997}. Although the formation of MP has been discussed for various different magnetic systems,
the underlying microscopic nature of electrical transport in the MP phases is not yet properly understood and is a matter of current debate \cite{MajumdarPRL1998, CalderonPRB2004, ChatterjeePRB2004, YuPRB2006}.

In EuB$_6$, magnetic phase separation has been suggested, e.g.\ from muon spin rotation experiments \cite{BrooksPRB2004}. However, direct evidence for electronic phase separation and percolation from transport measurements has been lacking so far.
Yet, an important hint came from recent Hall effect measurements by Zhang {\it et al.}, who observed a distinct change in the slope of the Hall resistivity in the PM phase \cite{ZhangPRL2008,ZhangPRL2009}. The switching field in the Hall effect depends linearly on temperature and extrapolates to the paramagnetic Curie temperature of the material.
The authors interpret the switching field occurring at a single critical magnetization as the point of percolation for patches of a more conducting and magnetically-ordered phase in a PM background. This picture yields excellent scaling in an empirical two-component model \cite{ZhangPRL2009}, providing a measure of the degree of electronic phase separation.

Motivated by these findings, we performed fluctuation spectroscopy and weak nonlinear transport (third-harmonic voltage) measurements, which --- unlike the conventional linear resistivity --- are sensitive to the microgeometry of the sample, i.e.\ intrinsic electronic inhomogeneities. We find direct evidence for electronic phase separation and magnetically-driven percolation, both when cooling through the FM transition and when applying magnetic fields in the PM temperature regime.

Single crystals of EuB$_6$ were grown from Al flux as described in \cite{FiskJAP1979}. Fluctuation spectroscopy measurements were carried out using a standard \emph{ac} method in a four-terminal setup \cite{ScofieldRSI1987,MuellerChemPhysChem2011}.
The (linear and nonlinear) transport measurements were carried out in standard four-probe geometry using {\it ac} lock-in technique at a frequency of 17\,Hz. For weakly nonlinear transport, current density and electrical field are related by $j = (\sigma  + b|E|^2) E$, where $\sigma$ is the Ohmic or linear conductivity and $b$ the nonlinear conductivity coefficient, with $b|E|^2 \ll \sigma$ \cite{BergmanPRB1989,Levy1994}. The cubic nonlinearity can be accessed in {\it ac} transport measurements by detecting a third-harmonic voltage signal $V_{3\omega}$, see e.g.\ \cite{BergmanPRB1989,DubsonPRB1989,MoshnyagaPRB2009}.
In general, noise spectroscopy and nonlinear transport are a measure of the fourth moment of the current distribution \cite{DubsonPRB1989} and therefore probe the microgeometry of the electronic system, which is not accessible by the linear (Ohmic) resistance, being related to the second moment of the current distribution \cite{Bergman1992}.

\begin{figure}[h]
\begin{center}
\includegraphics[width=0.475\textwidth]{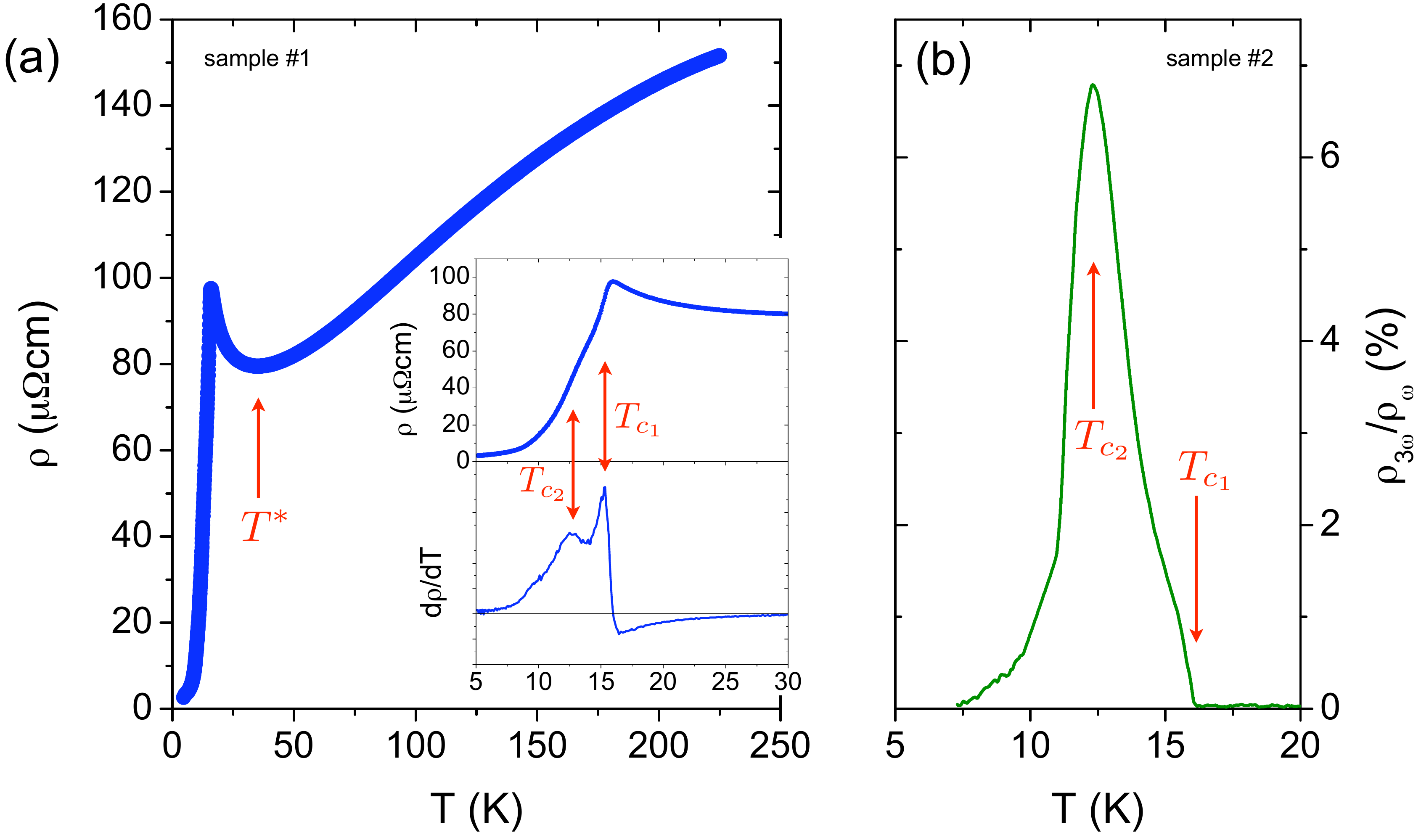}
\caption{\label{figure1} (Color online) (a) Resistivity of the EuB$_6$ sample (\#1) used for the noise measurements shown in Fig.\,\ref{figure2}. Inset: Enlarged low-temperature region. Peaks in the temperature derivative ${\rm d}\rho/{\rm d}T$ mark the two transitions at $T_{c_1} = 15.2$\,K and $T_{c_2} = 12.6$\,K. (b) The coefficient of weak nonlinear transport measured by third-harmonic voltage generation, $\kappa = \rho_{3 \omega}/\rho_\omega = V_{3 \omega}/V_\omega$, for sample \#2.}
\end{center}
\end{figure}
Figure\,\ref{figure1}(a) shows the bulk resistivity of a representative EuB$_6$ sample (\#1, optimized shape for noise measurements). Upon cooling, the resistivity decreases from room temperature down to $T^\ast \sim 35$\,K, where a broad minimum is observed. Below that temperature, the resistivity increases and goes through a maximum at about $\sim 16$\,K, before it rapidly decreases. As seen in the inset of Fig.\,\ref{figure1}(a), below this sharp drop there is a shoulder marking the second, lower transition. In the literature, the two transitions are often defined as pronounced peaks in the temperature derivative ${\rm d}\rho/{\rm d}T$ observed here at $T_{c_1} = 15.2 $\,K (just below the resistivity maximum) and $T_{c_2} = 12.6$\,K, see inset of Fig.\,\ref{figure1}(a). Such clearly visible transitions are observed only for high-quality samples \cite{SuellowPRB1998}.
Interestingly, $T_{c_1}$ and $T_{c_2}$ coincide with the onset and peak, respectively, of a weak nonlinear contribution to the electronic transport, shown in Fig.\,\ref{figure1}(b), measured by the ratio of third-harmonic and linear resistivity $\rho_{3\omega}/\rho_\omega$ shown here for sample \#2. Clearly, in this temperature region the electronic system exhibits substantial local inhomogeneities.
We note that although there is a slight sample-to-sample variation in $T_{c_1}$ and $T_{c_2}$ (16.1\,K and 12.4\,K, respectively for sample \#2, and 15\,K and 12.3\,K for \#3), the relation between the kink in resistance and the peak in the nonlinear signal is ubiquitous in the samples we have measured.

\begin{figure}[h]
\begin{center}
\includegraphics[width=0.485\textwidth]{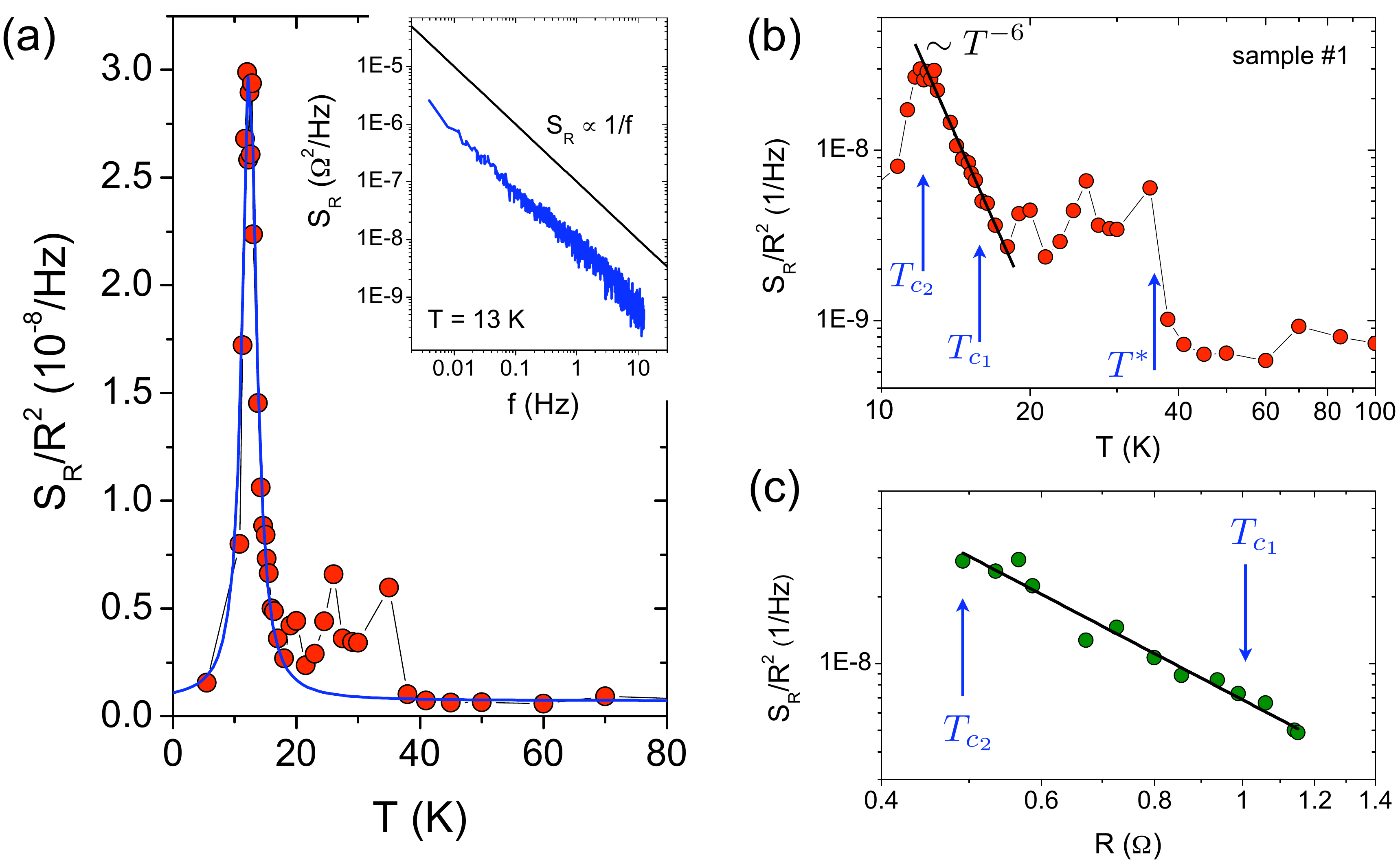}
\caption{\label{figure2} (Color online) (a) Temperature dependence of the normalized noise PSD $S_R/R^2(T, f = 1\,{\rm Hz})$. Inset exemplarily shows a typical spectrum obtained at $T = 13$\,K, being close to $S_R \propto 1/f$ (solid line). The solid line in the main panel shows a fit parameterizing a diverging behavior with a peak at $T = 12.4$\,K and a width $\Delta T\,=\,2.4$\,K. (b) Same data in a log-log plot highlighting a step-like increase of the $1/f$-noise level below $T^\ast \sim 35$\,K and a power-law divergence following roughly $S_R/R^2 \propto T^{-6}$ (solid line). (c) Scaling $S_R/R^2 \propto R^{w}$ in the percolation regime $T_{c_2} \leq T \sim T_{c_1}$. A fit to the data yields $w \sim - 2.1$ (solid line).}
\end{center}
\end{figure}
Next, we focus on the measurements of the resistance noise power spectral density (PSD), $S_R(f,T)$. 
A typical spectrum, measured at $T = 13$\,K, is shown in the inset of Fig.\,\ref{figure2}(a). In the entire temperature range we observe $1/f$-type noise, characterizing the intrinsic equilibrium conductance fluctuations of the sample. As required, this excess noise is quadratic in the current through the sample (not shown). The main panel of Fig.\,\ref{figure2}(a) shows the normalized resistance noise PSD $S_R/R^2(T)$ taken at 1\,Hz. The data can be described by a parameterized divergence with the peak at a temperature $T = 12.4$\,K coinciding with $T_{c_2}$ and a width $\Delta T = 2.4$\,K. Furthermore, two striking features in the noise measurements are (i) a step-like increase below about 35\,K coinciding with the minimum of the resistivity at $T^\ast$ (seen more clearly in Fig.\,\ref{figure2}(b)), and (ii) a power-law divergence starting above $T_{c_1}$ with a pronounced peak at $T_{c_2}$, below which the noise level rapidly drops again.
As discussed in detail below, we interpret this behavior as a reflection of the percolation
of the poorly conducting component of the sample, with the peak determining the percolation threshold at $T_{c_2}$.
Between $T_{c_2}$ and $T_{c_1}$, the noise scales as $S_R/R^2 \propto T^{\beta}$ with $\beta \sim - 6.5 \pm 0.6$, remarkably similar to the power law observed when approaching the metal-insulator transition in bulk samples of disordered P-doped Si \cite{KarPRL2003}.

In general, diverging $1/f$-type resistance (or conductance) fluctuations are typical for percolative metal-insulator transitions \cite{RammalPRL1985,Kogan1996}.
In a classical percolation scenario, the resistance and resistance noise, respectively, are expected to scale as $R$, $S_R/R^2 \propto (p-p_c)^{-t, \kappa}$. Here, $p$ is the fraction of unbroken bonds of a random resistor network, $p_c$ the percolation threshold, and $t$ and $\kappa$ the critical exponents for the resistivity and noise, respectively. Since the microscopic details determining $p$ are not accessible in most cases, it is convenient to express the normalized noise as a function of resistance $S_R/R^2 \propto R^{w}$,
with $w = \kappa/t$ \cite{YagilPRB1992}. The scaling exponents $t$, $\kappa$ and $w$ depend on the type of percolating system and its dimensionality and are determined by computer simulations, where exact solutions are not known \cite{Stauffer-Buch}.
The divergence of the noise close to $p_c$ usually is explained by the reduced number of effective current paths, which results in the suppression of cancellation of uncorrelated resistance fluctuations along different paths, which are abundant far away from $p_c$. This effect is similar to the inverse dependence of the noise PSD on the volume of macroscopic bodies \cite{HoogePLA1969}.
In a {\it composite} system, $p$ may be the portion of the metallic phase in a less conducting background, the portion of which is $q$. In certain cases, the percolation threshold $p_c$ of that metallic phase can be different from the percolation threshold $q_c$ of the less conducting or insulating portion \cite{Yu1994,Levy1994}. Therefore, depending on which contribution dominates, the percolation threshold probed by noise spectroscopy/nonlinear transport on the one hand and conventional linear resistivity on the other hand may be different. In EuB$_6$, we consider the MP as entities of a more conducting and magnetically-ordered phase in a paramagnetic and 'poorly conducting' background \cite{SuellowPRB2000,ZhangPRL2009}, which form links at $T_{c_1}$ corresponding to percolation threshold $p_c$ and therefore a continuous conduction path through the sample leading to delocalization of holes and hence the drop of the resistance. Indeed, S\"ullow {\it et al.}\ \cite{SuellowPRB2000} have ascribed $T_{c_1} = T_M$ to a metallization transition via the overlap of magnetic polarons.
The separation of the charge delocalization and bulk magnetic ordering at $T_{c_2} = T_C$ \cite{SuellowPRB2000} then implies electronic and magnetic phase separation.
The noise measurements clearly show a percolation threshold $q_c$ at $T_{c_2}$, which means that the resistance and its fluctuations are sensitive to different parts of the conducting network. Consequently, we observe a {\it negative} scaling exponent $w \sim -2.1 \pm 0.3$ above the peak of the noise, i.e.\ for $q > q_c$, where the resistivity increases with increasing temperature. Remarkably, such a clear separation is not a unique property of EuB$_6$ but also has been observed in perovskite manganites \cite{PodzorovPRB2000}, where, however, a two-component percolation scenario is not discussed.
Close to the threshold, the continuum 3D 'random void' and 'inverted random void' models have scaling exponents $w = 2.1$ and $2.4$, respectively \cite{TremblayPRB1986}. For the present data, the scaling on the low-temperature side, i.e.\ below the percolation threshold, could not be verified due to the limited number of data points.

As shown in Fig.\,\ref{figure1}\,(b), the characteristic temperatures $T_{c_1}$ and $T_{c_2}$ are clearly seen also in the weak nonlinear transport, as the onset and pronounced peak, respectively. In a percolative system close to the threshold, the local current densities and electric fields can be much larger than their average values. This is due to narrow paths, so-called 'bottlenecks' and 'hot spots' in the random network, the contribution of which to the resistance is proportional to the local electric field squared. Since $R(E) \approx R_\omega + AE^2 + \cdots$ with the second term being the third-harmonic nonlinear resistance, the magnitude of $R_{3\omega}$ indirectly reflects the density and local arrangement of the MP, see also \cite{MoshnyagaPRB2009}.
Therefore, we consider the electronic system as consisting of magnetically ordered spheres, which --- when they overlap --- form more conducting clusters, characterized by a conductivity $\sigma^{+}$, embedded in less conducting regions with $\sigma^{-}$. Both $S_R/R^2$ and $\rho_{3 \omega}/\rho_{\omega}$ peak at $T_{c_2}$, the percolation threshold $q_c$ for the less conducting regions, which is reached when the MP clusters merge and start forming a continuum. Most interestingly, $T_{c_2}$ (and $q_c$) coincides with the material's bulk Curie temperature as determined from magnetization measurements \cite{SuellowPRB2000}. This important observation
may be understood by taking into account that the zero-field magnetic susceptibility diverges at $T_C$, which is most favorable for the stabilization of MP. Hence, at this temperature, the density of MP should be maximum, which is the criterion for the percolation threshold we observe.
In theoretical studies it is found \cite{BergmanPRB1989} that weak nonlinear transport and $1/f$ noise behavior of a composite system of conducting spheres in a less conducting background, such as the present one, is more complex than for classical bond or site percolation. This is because of the role of the local microgeometry, parameterized by the channel width between the conducting spheres and the degree of their overlap, which in turn is related to substantially enhanced local electric fields. These local geometric parameters, the degree of disorder of the MP (i.e.\ deviations from a regular array) and the ratios $\sigma^{+}/\sigma^{-}$ and $S_{R}^+/S_{R}^-$ determine whether the good or the bad conductor dominates the overall noise and the nonlinear transport signal. In the region of the FM transition, obviously the contribution of the background electrons, i.e.\ the poor conductor, dominates.

\begin{figure}[h]
\begin{center}
\includegraphics[width=0.475\textwidth]{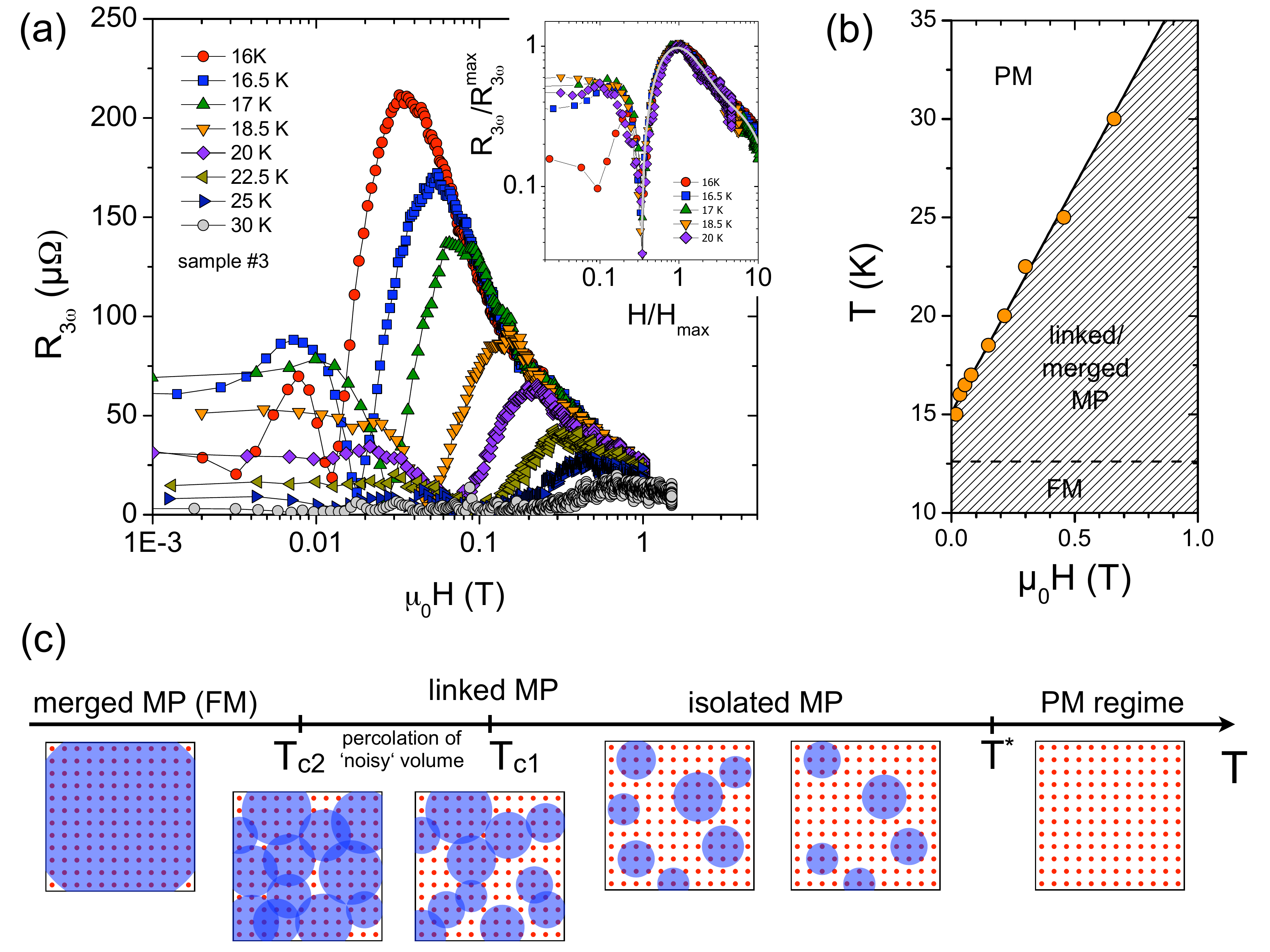}
\caption{(Color online) (a) $R_{3\omega}$ as a function of magnetic field $H$ measured at different temperatures in the PM regime. Inset shows the same data normalized to the field value and magnitude of the maximum in a log-log plot, demonstrating that the data collapse on one curve \cite{Note1}. (b) Phase diagram of EuB$_6$ determined from the field sweeps in (a). Circles denote the position of the maximum in $R_{3 \omega}(H)$ showing a linear behavior. 
(c) Schematic of the MP states at different temperatures in zero magnetic field (after \cite{YuPRB2006}). Note that the schematic is a 2D projection of spherical MP in 3D.
\label{figure3}}
\end{center}
\end{figure}
The MP are suggested to form (to become stabilized) below $T^\ast \sim 35$\,K \cite{NyhusPRB1997,SuellowPRB2000,BrooksPRB2004}. At about the same temperature, the noise shows a pronounced step-like increase upon cooling, which can be interpreted as hopping of magnetic clusters, the number of which increases as the temperature is lowered, and/or scattering by these magnetic excitations. The observed peak structure in $S_R/R^2(T)$ may reflect the energetics of such a mechanism. In the PM region $T_{c_1} < T < T^\ast$, the MP are initially isolated and diluted in the electronic 'background sea'.
When a magnetic field is applied, the MP increase in size \cite{Note2} until they form links, eventually overlap (where at the same time their number may decrease) and finally merge. Figure\,\ref{figure3}(a) shows a pronounced maximum in $R_{3\omega}(H)$ (corresponding to $q_c$) for different $T$ in the PM regime, which succeeds the field-induced percolation of the MP (corresponding to $p_c$), similar to the situation when cooling in zero magnetic field through the FM transition. Indeed, a drop of the sample resistance precedes the peak in $R_{3\omega}$ (not shown). A striking observation is a linear temperature dependence of $R_{3\omega}^{\rm max}(H)$, see Fig.\,\ref{figure3}(b). Thus, like the switching field in the Hall effect \cite{ZhangPRL2009},
the maximum in nonlinear transport occurs at a single critical magnetization.
The magnitude of $R_{3\omega}(T,H)$ is related to the microgeometry of the phase-separated electronic system reflecting the density and arrangement of the MP.
Accordingly, at high temperatures, slightly below $T^\ast$, only a few MP are stable and large fields are necessary in order to achieve a substantial overlap being accompanied by charge-carrier delocalization. Upon lowering the temperature, obviously the number density of MP increases and smaller fields are needed to reach the percolation threshold. Also, the larger magnitude of $R_{3\omega}$ indicates a higher degree of inhomogeneity due to the larger number of polaronic objects. Since the normalized measurements collapse onto one curve \cite{Note1}, a single function, see inset of Fig.\,\ref{figure3}(a), describes the density of magnetic polarons at different temperatures and magnetic fields.

Work is supported by the DFG through the Emmy Noether program and the NSF through DMR-0801253.


\begin{thebibliography}{99}

\bibitem{Dagotto2001} E. Dagotto, T. Hotta, and A. Moreo, Phys. Rep. \textbf{344}, 1 (2001).
\bibitem{DegiorgiPRL1997} L. Degiorgi, E. Felder,  H. R. Ott, J. L. Sarrao, and Z. Fisk, Phys. Rev. Lett. \textbf{79}, 5134 (1997).
\bibitem{CooleyPRB1997}J. C. Cooley, M. C. Aronson, J. L. Sarrao and Z. Fisk, Phys. Rev. B \textbf{56},14541 (1997).
\bibitem{MajumdarPRL1998} P. Majumdar and P. Littlewood, Phys. Rev. Lett. \textbf{81}, 1314 (1998).
\bibitem{TeresaNature1997} J. M. De Teresa, M. R. Ibarra, P. A. Igarabel, C. Ritter, C. Marquina, J. Blasco, J. Garcia, A. del Moral, and Z. Arnold, Nature \textbf{386}, 256 (1997).
\bibitem{NyhusPRB1997}P. Nyhus, S. Yoon, M. Kauffman, S. L. Cooper, Z. Fisk and J. Sarrao, Phys. Rev. B \textbf{56}, 2717 (1997).
\bibitem{SuellowPRB2000} S. S\"ullow, I. Prasad, M. C. Aronson, S. Bogdanovich, J. L. Sarrao, and Z. Fisk,  Phys. Rev. B \textbf{62}, 11626 (2000).
\bibitem{Molnar1967} S. von Moln\'{a}r and S. Methfessel, J. Appl. Phys. {\bf 38}, 959 (1967).
\bibitem{KasuyaRMP1968} T. Kasuya and A. Yanase, Rev. Mod. Phys. \textbf{40}, 684 (1968).
\bibitem{vonMolnarHandbook} S. von Moln\'{a}r and P. A. Stampe, Hand book of Magnetism and Advanced Magnetic Materials, Vol 5, Wiley Publishers.
\bibitem{CalderonPRB2004} M. J. Calder\'{o}n, L. G. L. Wegener, and P. B. Littlewood, Phys. Rev. B \textbf{70}, 092408 (2004).
\bibitem{ChatterjeePRB2004} J. Chatterjee, U. Yu, and B. Min, Phys. Rev. B \textbf{69}, 134423 (2004).
\bibitem{YuPRB2006} U. Yu and B. I. Min, Phys. Rev. B \textbf{74}, 094413 (2006).
\bibitem{BrooksPRB2004} M. L. Brooks, T. Lancaster, S. J. Blundell, W. Hayes, F. L. Pratt, and Z. Fisk, Phys. Rev. B \textbf{70}, 020401(R), (2004).
\bibitem{ZhangPRL2008} X. Zhang, S. von Moln\'{a}r, Z. Fisk, and P. Xiong, Phys. Rev. Lett. \textbf{100}, 167001 (2008).
\bibitem{ZhangPRL2009}X. Zhang, L. Yu, S. von Moln\'{a}r, Z. Fisk and P. Xiong, Phys. Rev. Lett. \textbf{103}, 106602 (2009).
\bibitem{FiskJAP1979}Z. Fisk \textit{et al.} J. Appl. Phys. \textbf{50}, 1911 (1979).
\bibitem{ScofieldRSI1987} J. H. Scofield, Rev. Sci. Instrum. \textbf{58}, 985 (1987).
\bibitem{MuellerChemPhysChem2011} J. M\"uller, ChemPhysChem {\bf 2011}, 12, 1222 -- 1245.
\bibitem{BergmanPRB1989} D. J. Bergman, Phys. Rev. B {\bf 39}, 4598 (1989).
\bibitem{Levy1994} O. Levy and D. J. Bergman, Phys. Rev. B {\bf 50}, 3652 (1994).
\bibitem{DubsonPRB1989} M. A. Dubson, Y. C. Hui, M. B. Weissman, and J. C. Garland, Phys. Rev. B \textbf{39}, 6807 (1989).
\bibitem{MoshnyagaPRB2009} V. Moshnyaga, K. Gehrke, O. I. Lebedev, L. Sudheendra, A. Belenchuk, S. Raabe, O. Shapoval, J. Verbeeck, G. Van Tendeloo, and K. Samwar, Phys. Rev. B \textbf{79}, 134413 (2009).
\bibitem{Bergman1992} D. J. Bergman and S. Stroud, {\em Physical properties of macroscopically inhomogeneous media}, in: Solid State Physics {\bf 46}, 147 -- 269 (1992).
\bibitem{SuellowPRB1998} S. S\"ullow, I. Prasad, M. C. Aronson, J. L. Sarrao, Z. Fisk, D. Hristova, A. H. Lacerda, M. F. Hundley, A. Vigilante and D. Gibbs, Phys. Rev. B \textbf{57}, 5860 (1998).
\bibitem{KarPRL2003} S. Kar, A. K. Raychaudhui, A. Ghosh, H. v. L\"ohneeysen, and G. Weiss, Phys. Rev. Lett. \textbf{91}, 216603 (2003).
\bibitem{RammalPRL1985} R. Rammal, C. Tannous, P. Breton, and A.-M. S. Trembley, Phys. Rev. Lett. \textbf{54}, 1718 (1985).
\bibitem{Kogan1996} Sh. Kogan, {\em Electronic noise and fluctuations in solids}, Cambridge University Press, New York (1996).
\bibitem{YagilPRB1992} Y. Yagil and G. Deutscher, Phys. Rev. B. \textbf{46}, 16115 (1992).
\bibitem{Stauffer-Buch}D. Stauffer and A. Aharony, {\em Introduction to percolation theory}, 2nd ed., Taylor \& Francis, London, Washington, DC (1992).
\bibitem{HoogePLA1969} F. N. Hooge, Phys. Lett. \textbf{29A}, 139 (1969).
\bibitem{Yu1994}K.W. Yu and P.M. Hui, Phys. Rev. B \textbf{50}, 13327 (1994).
\bibitem{PodzorovPRB2000} V. Podzorov, M. Uehara, M. E. Gershenson, T. Y. Koo, and S-W. Cheong, Phys Rev B \textbf{61}, R3784 (2000).
\bibitem{TremblayPRB1986}A.-M. S. Tremblay, S. Feng, and P. Breton, Phys. Rev. B \textbf{33}, 2077 (1986).
\bibitem{Note1} The behavior at low fields is not fully understood but seems to be related to the near vicinity to the magnetic ordering temperature.
\bibitem{Note2} Spatial extent of magnetic polarons were found to be increased by application of external magnetic fields from the observation of ferromagnetic correlation length in manganites \cite{TeresaNature1997}. The data also suggest a reduction of the number of polarons due to application of magnetic field.





\end{thebibliography}
\end{document}